\documentclass[12pt,preprint]{aastex}

\begin{document}

\title{Evolution of the high-mass end of the  stellar initial mass functions
in starburst galaxies}

\author{Kenji Bekki and  Gerhardt R. Meurer} 
\affil{
ICRAR,
M468,
The University of Western Australia
35 Stirling Highway, Crawley
Western Australia, 6009, Australia
}

\begin{abstract}
We investigate the time evolution and spatial variation of
the stellar initial mass function (IMF) in star-forming disk galaxies
by using chemodynamical simulations with an IMF model depending both on local 
densities and metallicities ([Fe/H]) of the interstellar medium
(ISM).
We find that the slope ($\alpha$)
of a power-law IMF ($N(m) \propto m^{-\alpha}$)
for stellar masses larger than $1M_{\odot}$
evolves from
the canonical Salpeter IMF ($\alpha \approx 2.35$)
to  be moderately top-heavy one ($\alpha \approx 1.9$)
in the simulated disk galaxies with starbursts triggered by galaxy interaction.
We also find that $\alpha$ in  star-forming regions
correlates 
with star formation rate densities ($\Sigma_{\rm SFR}$
in units of $M_{\odot}$ yr$^{-1}$ kpc$^{-2}$).
Feedback effects of Type Ia and II supernovae are found to prevent IMFs from being
too top-heavy ($\alpha < 1.5$).
The simulation predicts 
$\alpha \approx  0.23 \log \Sigma_{\rm SFR} + 1.7$ 
for $\log \Sigma_{\rm SFR} \ge -2$ 
(i.e.,  more top-heavy in
higher $\Sigma_{\rm SFR}$), which is reasonably consistent  well with
corresponding recent observational results.
The present study also predicts that 
inner regions of starburst disk galaxies have 
smaller $\alpha$ thus are more top-heavy ($d\alpha/dR \sim 0.07$ kpc$^{-1}$ for $R\le 5$ kpc).
The predicted radial $\alpha$ gradient can be tested against future observational
studies of the  $\alpha$ variation in star-forming galaxies.
\end{abstract}

\section{Introduction}

A growing number of recent observational studies have suggested
that stellar initial mass functions (IMFs) can be different in galaxies
with different star-formation activities,  dynamical parameters
(e.g., stellar velocity dispersion), environments, 
and redshifts (e.g.,  Hoversten \& Glazebrook 2008;
van Dokkum 2008; Meurer et al. 2009, M09; Treu et al. 2010;
Cappellari et al. 2012;
Ferreras et al. 2012; van Dokkum \& Conroy 2012).  
One of important questions relating to  possibly variable IMFs 
is as to whether IMFs can be top-heavy in starburst
galaxies (e.g., Elmegreen 2009 for a recent review).
It has been still controversial whether 
the IMF is almost universal in different galaxies or depends on local environments 
such as densities and metallicities of star-forming gas clouds
(e.g., Bastian et al. 2010;
Kroupa et al. 2012).

Recent observational studies on the physical relations between the
slopes of power-law IMFs ($\alpha$,
where $\alpha=2.35$ is the canonical Salpeter IMF)
and galaxy properties (e.g., mass densities and SFRs) have provided
new clues to the nature of IMFs in actively star-forming  galaxies.
M09 investigated the extinction-corrected flux ratio
of ${\rm H}_{\alpha}$ and far-ultraviolet (FUV) light and found that the flux ratios 
are different between different galaxies and depend on surface densities
of ${\rm H}_{\alpha}$ flux ($\Sigma_{\rm H_{\alpha}}$)
and $R$-band luminosity densities of galaxies ($\Sigma_{\rm R}$).
Their observations are consistent 
with $\alpha$ decreasing  (i.e., becoming more top-heavy)
with increasing  $\Sigma_{\rm H_{\alpha}}$  and $\Sigma_{\rm R}$.  
Gunawardhana et al. (2011, G11) investigated $\alpha$ of $\sim 33000$ galaxies
and their correlations with SFRs (in units of $M_{\odot}$ yr$^{-1}$) 
and SFR densities (in units of $M_{\odot}$ yr$^{-1}$ kpc$^{-2}$)
and found correlations between $\alpha$ and SFRs 
($\alpha \approx -0.36 \log \langle {\rm SFR} \rangle + 2.6$) and between
$\alpha$ and $\Sigma_{\rm SFR}$ 
($\alpha \approx -0.3 \log \langle \Sigma_{\rm SFR} \rangle +1.7$).
It should be noted here that 
they  take a linear scaling between H$\alpha$ luminosity and the SFR 
based on $\alpha = 2.35$  in deriving the SFR.
In reality, a linear scaling does not apply 
if $\alpha$ is varying: the exact scaling dependent on 
the behavior of the IMF in detail (e.g. the mass limits and 
any changes in $\alpha$ at the low mass end)
is beyond the scope of this paper to determine. 
Although these observed correlations appear to strongly suggest
top-heavy IMFs in starburst galaxies, 
the origin of the correlations has been 
discussed by only a few theoretical models 
in a quantitative manner 
(Pflamm-Altenburg et al. 2009; Weidner et al. 2011; Kroupa et al. 2012).

Pipino \&  Matteucci (2004) showed that the observed colors of massive elliptical
galaxies can be reproduced better by their chemical and photometric evolution
models with slightly flatter (i.e., top-heavy) IMFs. Nagashima et al. (2005) demonstrated that
the observed chemical abundances (e.g., [Mg/Fe])
of elliptical galaxies can be explained by
top-heavy IMFs. These previous works, however,  did not investigate how IMFs
evolve in the starburst phases of elliptical galaxy formation.
Although starbursts have been demonstrated to be triggered by galaxy interaction
in previous numerical simulations (e.g., Noguchi \& Ishibashi 1986), the evolution
of IMFs during  starbursts has not been investigated in the simulations
of galaxy interaction.
Thus it is theoretically unclear whether or not the IMFs of galaxies
become top-heavy when they are in the starburst phases.

The two main purposes of this {\it Letter} are as follows. 
Firstly, we discuss whether or not the  IMFs in starburst galaxies can become top-heavy
by using our original chemodynamical simulations that incorporate the new physical
IMF model proposed by Marks et al. (2012, M12). 
Secondly, we test whether or not the present model
can reproduce the observed correlations between $\alpha$, SFRs, and $\Sigma_{\rm SFR}$
derived by G11.
We investigate $\alpha$ at individual star-forming regions
in disk galaxies
at each time step
so that we can investigate not only the time evolution of IMF 
but also local variation of $\alpha$. 
We focus our  investigation on the 
 IMF evolution of galaxies with starbursts triggered
by strong galaxy-galaxy interaction and merging.
In the present paper, we compare between
 the observed and simulated $\alpha-\Sigma_{\rm SFR}$
relations, even though the observed relation would be less precisely derived
by assuming the fixed Salpter IMF for the estimation of SFRs.

\section{The model}
In order to perform  numerical simulations of galaxies  on GPU clusters,
we use our original chemodynamical code (Bekki 2013) 
that is a revised version of our previous  code (`GRAPE-SPH'; Bekki 2009).
The new GPU-version chemodynamical code allows us to  investigate 
the time evolution of chemical abundances and dust properties of galaxies
self-consistently. We adopt a disk galaxy model that broadly mimics the structural 
and kinematical properties of baryonic and dark matter components
of the Milky Way Galaxy (as listed in Binney \& Tremaine 2007).
The disk galaxy is composed of dark matter halo, bulge, stellar disk, and
gas disk. The density distribution of the dark halo is represented by the `NFW' profile
predicted by the cold dark matter cosmology
(Navarro et al. 1996),
and the total mass, the virial radius, and the $c$-parameter are set to be
$10^{12} M_{\odot}$, 245 kpc, and 10, respectively.

The stellar bulge is assumed to have the Hernquist profile with the total mass
of $10^{10} M_{\odot}$ and the disk size of 3.5 kpc. 
The radial ($R$) and vertical ($Z$) density profiles of the stellar disk are
assumed to be proportional to $\exp (-R/R_{0}) $ with scale
length $R_{0}$ = 3.5 kpc and to  ${\rm sech}^2 (Z/Z_{0})$ with scale
height  $Z_{0}$ = 0.7 kpc, respectively.
In addition to the
rotational velocity caused by the gravitational field of disk,
bulge, and dark halo components, the initial radial and azimuthal
velocity dispersions are assigned to the disc component according to
the epicyclic theory with Toomre's parameter $Q$ = 1.5.
The maximum circular velocity of the disk is 222 km s$^{-1}$ for the adopted
structure parameters of the stellar and dark matter components.

The  gas disk represented by SPH particles has a  gas mass fraction of 0.1 and
an exponential radial profile with the scale-length of 3.5  kpc 
and the size of 35 kpc. 
The gas disk has an initial temperature of $10^4$ K and a metallicity gradient 
with the central [Fe/H] of 0.34 and the slope ($d{\rm [Fe/H]}/dR$) of $-0.04$
dex kpc$^{-1}$ (Andrievsky et al. 2004).
The gas particles are converted into new stars if the following physical conditions
are met:
(i) the local dynamical time scale is shorter
than the sound crossing time scale (mimicking
the Jeans instability)  and (ii) the local velocity
field is identified as being consistent with gravitationally collapsing
(i.e., div {\bf v}$<0$). 
The feedback energy of $10^{51}$ erg is imparted to neighboring
gas particles around a new star  when the star becomes a Type Ia or Type II supernova (SN Ia and SN II,
respectively).

For a new star formed from Jeans-unstable gas, the IMF slope ($\alpha$) 
for stellar masses ($m_{\rm s}$) larger
than $1 M_{\odot}$ 
is investigated.  
In the following, the high-mass IMF slope (originally $\alpha_3$ in 
the Kroupa IMF; Kroupa 2001) 
is referred to as $\alpha$ for convenience. 
M12 proposed that $\alpha$  
depends on mass densities ($\rho_{\rm cl}$)
and [Fe/H] of proto-cluster gas clouds
as follows:
\begin{equation}
\alpha = 0.0572 \times {\rm [Fe/H]} -0.4072 \times \log_{10}(
\frac{ \rho_{\rm cl} }{ 10^6 M_{\odot} {\rm pc}^{-3} }) +1.9283 
\end{equation}
This equation holds for $x_{\rm th} \ge -0.87$, where
$x_{\rm th}=-0.1405{\rm [Fe/H]}+\log_{10}(
\frac{ \rho_{\rm cl} }{ 10^6 M_{\odot} {\rm pc}^{-3} })$,
and $\alpha=2.3$ for $x_{\rm th} <-0.87$ (M12).  We slightly modify
the M12's IMF model in the following two points.
Firstly,  we do not introduce the threshold $x_{\rm th}$,
mainly because the $\alpha$-dependence at $x < x_{\rm th}$ is not
so clear (not so flat as M12 showed) owing to a small number of data points.

Secondly, we estimate  $\rho_{\rm cl}$ from local gas density $\rho_{\rm g}$
of Jeans-unstable gas particles
in  equation (1).  This is because the present simulation can not resolve
the rather high-density cores of star-forming or cluster-forming
molecular gas clouds. 
The local value of $\rho_{\rm cl}$ at each Jeans-unstable gas is
estimated by multiplying $\rho_{\rm g}$ by a constant $k_{\rm g}$:
\begin{equation}
\rho_{\rm cl}=k_{\rm g}\rho_{\rm g}.
\end{equation}
The derivation process of $k_{\rm g}$ is as follows. 
The Jeans-unstable gas is assumed to be star-forming giant molecular
clouds and thus have the following  size-mass scaling relation 
derived from the observed mass-density relation by Larson (1981):
\begin{equation}
R_{\rm gmc}=40 \times (\frac{ M_{\rm gmc} }{ 5\times 10^5 M_{\odot} })^{0.53} {\rm pc}.
\end{equation}
Since the equation  (1) is based largely on the observed properties of the Galactic 
globular clusters (GCs hereafter)
and nearby star clusters,
a reasonable  $k_{\rm g}$ can be the typical density ratio of GCs to GC-host GMCs.
We here  consider that (i) $\rho_{\rm cl}$ should correspond to a typical mean mass
density for GCs,
(ii) typical GC mass ($M_{\rm gc}$) and size ($R_{\rm gc}$) are $2 \times 10^5 M_{\odot}$
and 3pc, respectively (Binney \& Tremaine 2007),
(iii) original GCs just after  their formation from GMCs
should be $\sim 10$ times more massive than
the present ones  (e.g., Decressin et al. 2010; Marks \& Kroupa 2010; Bekki 2011),
and (iv) a star formation efficiency of GC-host
GMCs is $\sim 0.1$.
For $M_{\rm gmc}=2\times 10^7 M_{\odot}$ and $R_{\rm gmc}=283$ pc 
in the typical GC-host GMC,
a reasonable $k_{\rm g}$ ($=(R_{\rm gmc}/R_{\rm gc})^{3}$)
is estimated to be  $8.4 \times 10^5$.  
The present models do not show large $\alpha$ ($\sim 3$; very steep IMF) that
is needed to explain the observed  $H_{\alpha}/$FUV flux ratios
of low surface brightness galaxies (LSBs)  in M09.
This  is mainly because star formation can occur only in higher density gaseous regions
where $\alpha$ can be smaller in the present models.
Although the original IMF model by M12 is theoretically derived from observations 
on physical properties of GCs and star clusters,
we assume that the IMF model
applies for all new  stars in each bin (i.e., not just for star clusters).

The disk galaxy is assumed to interact with a companion galaxy represented by a point-mass
particle. The mass-ratio of the companion to the disk galaxy is a free parameter
denoted as $m_2$. The initial distance of the two galaxies is fixed
at 280 kpc and the orbital eccentricity ($e$)
and the orbital pericenter ($r_{\rm p}$) are free parameters.
The  angle between the $z$ axis and the vector of
the angular momentum of the disk is denoted as
$\theta$.
We mainly describe the results of 
the fiducial model with $m_2=1$, $e=1$ (i.e., parabolic), 
$r_{\rm p}=35$ kpc,
and $\theta=0^{\circ}$ (i.e., prograde interaction).
In order to discuss briefly the parameter dependences,
we investigate the following five representative  models:
weaker ($m_2=0.3$) and stronger ($m_2=3$) tidal interaction,
retrograde interaction ($\theta=180^{\circ}$),
low surface brightness (LSB; $R_0=5.5$ kpc), and no SN feedback effects.
For comparison, we also investigate the isolated model (i.e., no tidal interaction)
and the major merger one in which two identical disks merge with each other in
a prograde-prograde manner.
For the merger model, $r_{\rm p}=17.5$ kpc and $e=0.8$ are adopted.
The total number of particles is 716700 for isolated and interaction models
and 1433400 for the merger model. The gravitational softening length is
2.3 kpc for dark matter and 0.25 kpc for baryonic components. 
In the following,  $T$ in a simulation  represents the time that has elapsed since
the simulation started.

\section{Results}

Figure 1 shows how $\alpha$ changes during a starburst triggered by galaxy
interaction in the fiducial model. After the pericenter passage of the companion
galaxy,  the strong tidal field can compress the gas disk so that rather 
high-density gaseous regions
can be formed.  As a result of this,  the star formation rate (SFR) of the disk galaxy
can be significantly enhanced (up to $\sim 10 M_{\odot}$) and the mean $\alpha$
($\bar{\alpha}$) increases from $\sim 2.2$ (pre-starburst) to $\sim 1.9$ (during starburst).
The time evolution of SFR is clearly synchronized with that of $\alpha$,
which implies that $\alpha$ can be described as a function of SFR in galaxies.
Individual local regions of the disk galaxy during the
triggered  starburst have different
$\alpha$ ranging from $\sim 1.5$ to $\sim 2.6$ owing to locally different
gas densities and [Fe/H].
These results thus demonstrate that IMFs can become moderately top-heavy
in starburst galaxies  because of the formation of rather high-density and Jeans-unstable
gaseous regions.  

Figure 2 shows that $\alpha$ in individual radial bins 
(i.e., azimuthally averaged $\alpha$) of the disk galaxy
correlates with SFR densities ($\Sigma_{\rm SFR}$) in the fiducial model. 
Although dispersions are large at a given $\Sigma_{\rm SFR}$, 
radial bins  with higher  $\Sigma_{\rm SFR}$ 
are more likely to show smaller $\alpha$ (i.e., more top-heavy IMFs).
These simulated positive correlations between $\alpha$ and $\Sigma_{\rm SFR}$
are qualitatively consistent with the corresponding observational results by G11.
Figure 3 describes the radial $\alpha$ gradients of the star-forming disk
at pre-starburst (T=0.3 Gyr), strong starburst (T=1.1 Gyr),
and post-starburst phase (T=1.7 Gyr).
Clearly, more top-heavy IMFs (i.e., smaller $\alpha$) 
can be seen in the inner regions of the interacting
disk galaxy for the three epochs.
The disk has $d\alpha /dR$ of $0.07$ kpc$^{-1}$ for the central 5 kpc
at $T=1.1$ and 1.7 Gyr.
The derived positive $\alpha$ gradients can be seen in all models of the present
study and thus regarded as a robust prediction.

Figure 4 shows the $\alpha-\Sigma_{\rm SFR}$ correlation of individual radial bins 
with $\log \Sigma_{\rm SFR} \ge -2$ 
in the five different interacting galaxies with starbursts. Clearly,
radial bins  with higher $\Sigma_{\rm SFR}$ can have more top-heavy IMFs 
(i.e., smaller $\alpha$). A least-squares fit to these simulated star-forming
regions gives
\begin{equation}
\alpha=-0.23\log \Sigma_{\rm SFR}+1.7 ,
\end{equation}
which is similar to the observed relation of 
$\alpha \approx -0.3\log\Sigma_{\rm SFR}+1.7$ in G11.
It should be noted here, however, that the observed relation
is derived from  {\it averaged} $\alpha$ and $\Sigma_{\rm SFR}$ 
(over individual galaxies)
whereas the simulated points are  derived from
the average of star forming regions in each radial bin,
with the points belonging to
different galaxies. 
It should be also noted that the simulated $\alpha-\Sigma_{\rm SFR}$ relations
can be slightly different between different models. For example, 
$\alpha=-0.28\log \Sigma_{\rm SFR}+1.7$ in the fiducial model, which is
pretty close to the observed relation  by G11,
whereas 
$\alpha=-0.25\log \Sigma_{\rm SFR}+1.7$ in the model with $m_2=0.3$.
The LSB model can show more top-heavy IMFs in some regions, because
high-density gaseous regions can form owing to efficient 
gas-transfer to the central
region during tidal interaction hence original LSB galaxies
achieves a high surface brightness in the interaction.

Figure 5 shows that (i) number distributions of $\alpha$ 
in the fiducial model and the model without SN feedback effects
are bimodal (i.e., two peaks) and (ii) they
are significantly different between the two models.
The first  peak at lower $\alpha$ (`low-$\alpha$ peak') is
due to a starburst and the second peak at higher $\alpha$ (`high-$\alpha$ peak')
represents the mean $\alpha$ for pre-starburst IMFs.
The model without SN feedback has the low-$\alpha$ peak at lower $\alpha$ 
($\alpha \sim 1.4$),
which means that the IMF is more top-heavy in comparison with the fiducial model.
This difference between the two models
clearly demonstrates that the feedback effects of SN Ia and SN II
can suppress the formation of overly high-density, Jeans-unstable gaseous regions
and therefore prevent IMFs from being too top-heavy ($\alpha <1.5$). 
Figure 5 also shows that the location of the low-$\alpha$ peak in the merger
model  is coincident
with that of the fiducial one, which confirms that the IMF can become moderately
top-heavy in major galaxy merging.

\section{Discussion and conclusions}

The consistency 
between the observed and simulated $\alpha-\Sigma_{\rm SFR}$ relations
implies that the adopted IMF model (M12) can be useful in simulating the time evolution
and spatial variation of $\alpha$ in galaxies.  However, this does not necessarily
mean that the adopted IMF model 
for Jeans-unstable gas clouds is the best one that can be adopted for
discussing galaxy evolution caused by varying IMFs.
Larson (1998) adopted an IMF that is similar to the Salpeter IMF at the upper end
and flatter one below a characteristic stellar mass ($m_{\rm ch}$,
which is typically $\sim 1M_{\odot}$ in the present day universe)
and discussed a number of observational results in terms of time-varying $m_{\rm ch}$.
Following the IMF model proposed by Larson (1998),
Dav\'e (2008) and Narayanan \& Dav\'e (2012)
considered IMFs with $m_{\rm ch}$ depending on redshifts
and thereby discussed the origin of the observed redshift-evolution of the
relation between
stellar masses and SFRs in galaxies.
It should be noted here that the IMF model by M12 is observationally
supported whereas the $m_{\rm ch}$ variation model  by Larson is a working hypothesis
for which there is not much observational evidence.

Given that a growing number of observations have recently revealed
possible evidence for varying IMFs (e.g., Kroupa et al. 2012),
it would be important for theoretical studies of galaxy formation and evolution
to incorporate varying IMFs in their models.
It is however unclear what varying IMF models should be adopted for galaxy formation
and evolution studies.  The results of numerical simulations on galaxy formation
would depend on whether IMF models with varying $m_{\rm ch}$ (with fixed $\alpha$) 
or those with
varying $\alpha$ are adopted.
A reasonable varying IMF model would need to explain recent observational results
on IMFs in nearby galaxies
such as correlations between $H_{\alpha}$/FUV flux ratios and galaxy properties
(e.g., M09) and between $H_{\alpha}$ equivalent width and optical colors 
(e.g., Hoversten \& Glazebrook 2008; G11)
in a self-consistent manner.

One  robust prediction in the present study is that inner regions of actively
star-forming disk galaxies have smaller $\alpha$ (i.e., more top-heavy). 
Such radial gradients of IMFs are also discussed in the context
of the integrated galaxy IMF theory
(Pflamm-Altenburg \& Kroupa 2008).
The predicted {\it positive} $\alpha$ gradient is a natural consequence of
the adopted IMF model: 
the metallicity gradient act to flatten the IMF with
increasing radius, but the $\rho_{\rm g}$-dependence is stronger so the new gradient is
a steeper IMF at larger radius.
 Since observational studies have not yet extensively investigated
radial (or azimuthal) gradients of IMFs in nearby galaxies,
it is not clear whether the predicted $\alpha$ gradient is qualitatively or 
consistent with observations. 
We suggest that observed radial gradients of IMFs and their correlations with
local galaxy properties (e.g., local gas densities
and $R$-band local surface luminosity densities) would give some constraints
on whether IMF models with varying $\alpha$ or $m_{\rm ch}$ should be adopted
for galaxy formation studies.

Recent observations have suggested that (i) massive early-type galaxies have 
the steeper (i.e., bottom-heavy)
IMF in the mass range $0.1 M_{\odot}$ to $1 M_{\odot}$ (e.g., 
van Dokkum \& Conroy 2011)
and (ii) the IMF slope depends on stellar velocity dispersions of the galaxies
(e.g., Ferreras et al. 2012).
Since metal-rich star formation can lead to ``bottom-heavy'' IMFs
for $m_{\rm s} < 1 M_{\odot}$  in M12,
it is our future study to incorporate an IMF for $m_{\rm s} < 1 M_{\odot}$
that depends on  physical properties of ISM in our chemodynamical simulations
for elliptical galaxy formation.
Such  more sophisticated chemodynamical simulations with variable IMF slopes
below and above $m_{\rm s}=1 M_{\odot}$ will enable us to discuss the origin of
the observed possibly bottom-heavy IMFs in massive early-type galaxies
(e.g., Cenarro et al. 2003).

\acknowledgments
We are grateful to the anonymous referee for constructive and
useful comments.
K.B. acknowledges the financial support of the Australian Research
Council throughout the course of this work.
Numerical computations reported here were carried out both on the GPU clusters
(Pleiades and Fornax)
at the University of Western Australia.

\begin{figure}
\plotone{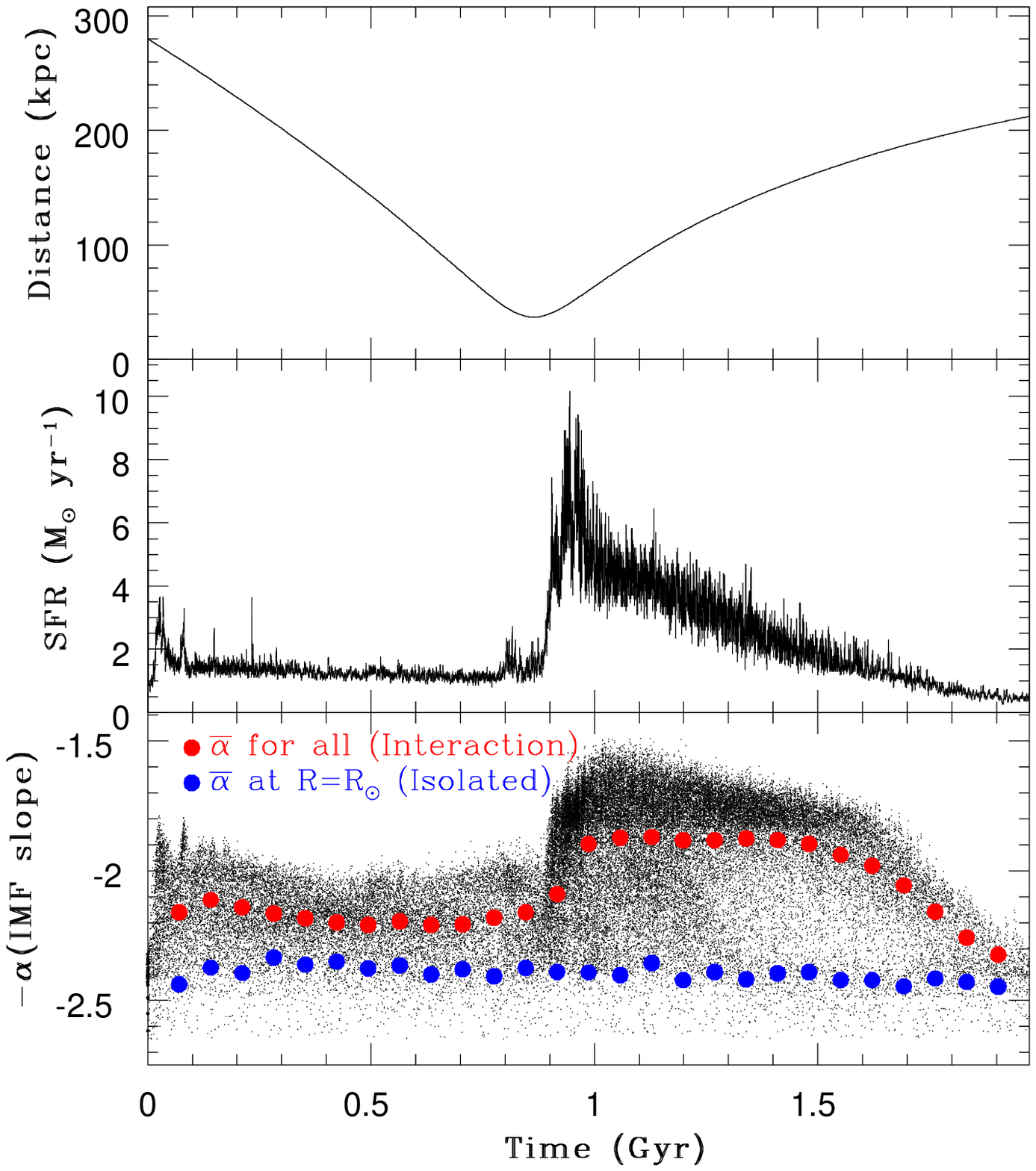}
\figcaption{
Time evolution of the distance of two interacting galaxies (top),
the star formation rate (SFR; middle), and the IMF slope $\alpha$ (bottom)
in the fiducial model. The big red and blue filled circles are the mean
$\alpha$ for all new stars at each time interval in the fiducial (interaction)
model and for those at the solar neighborhood ($6 \le R \le 10$ kpc) in the isolated
model, respectively.
The solar neighborhood in the isolated model has almost constant 
$\alpha$ ($\sim 2.3-2.4$; close to the Salpeter IMF), which strongly
suggests that the adopted varying IMF model does a good job in predicting
the IMF evolution.
\label{fig-1}}
\end{figure}

\begin{figure}
\plotone{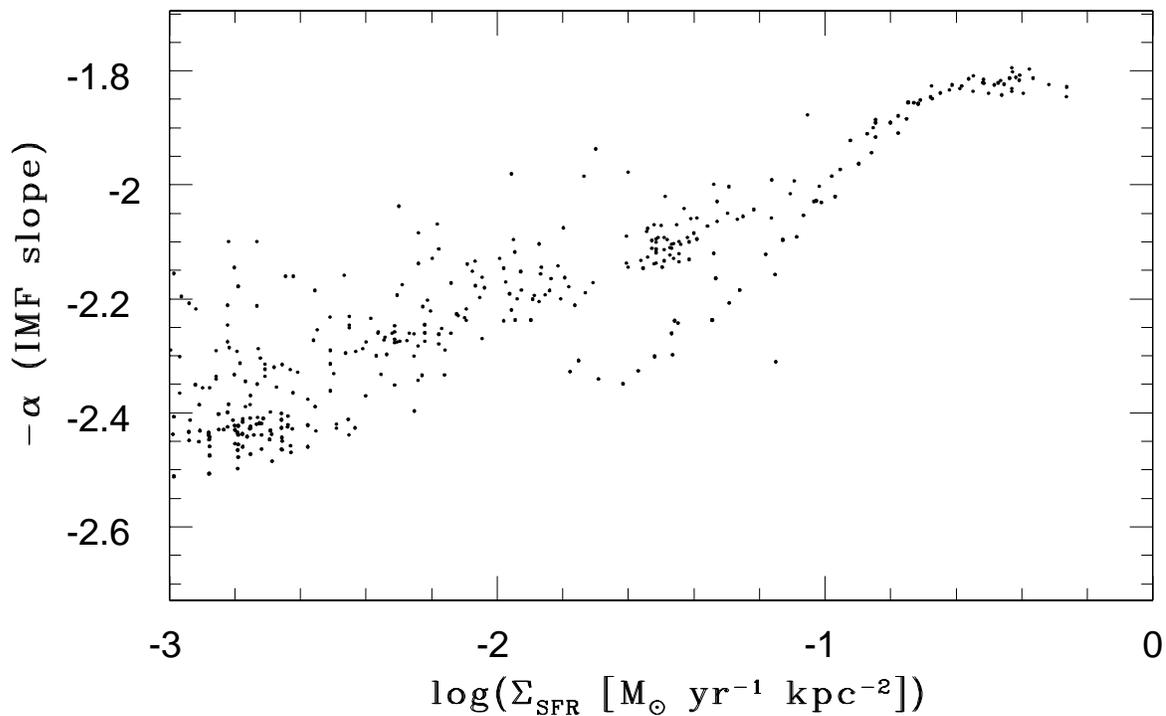}
\figcaption{
The locations of  star-forming regions 
on the $\log \Sigma_{\rm SFR}-\alpha$ plane
in the fiducial model.  For convenience,
$-\alpha$ is plotted against $\log \Sigma_{\rm SFR}$.
$\Sigma_{\rm SFR}$ are estimated at 10 radial bins ($R \le 17.5$ kpc) 
for star-forming regions 
(i.e., for only  new stellar particles with ages less than $\sim 10^7$ yr)
at each time step of the simulation.
\label{fig-2}}
\end{figure}

\begin{figure}
\plotone{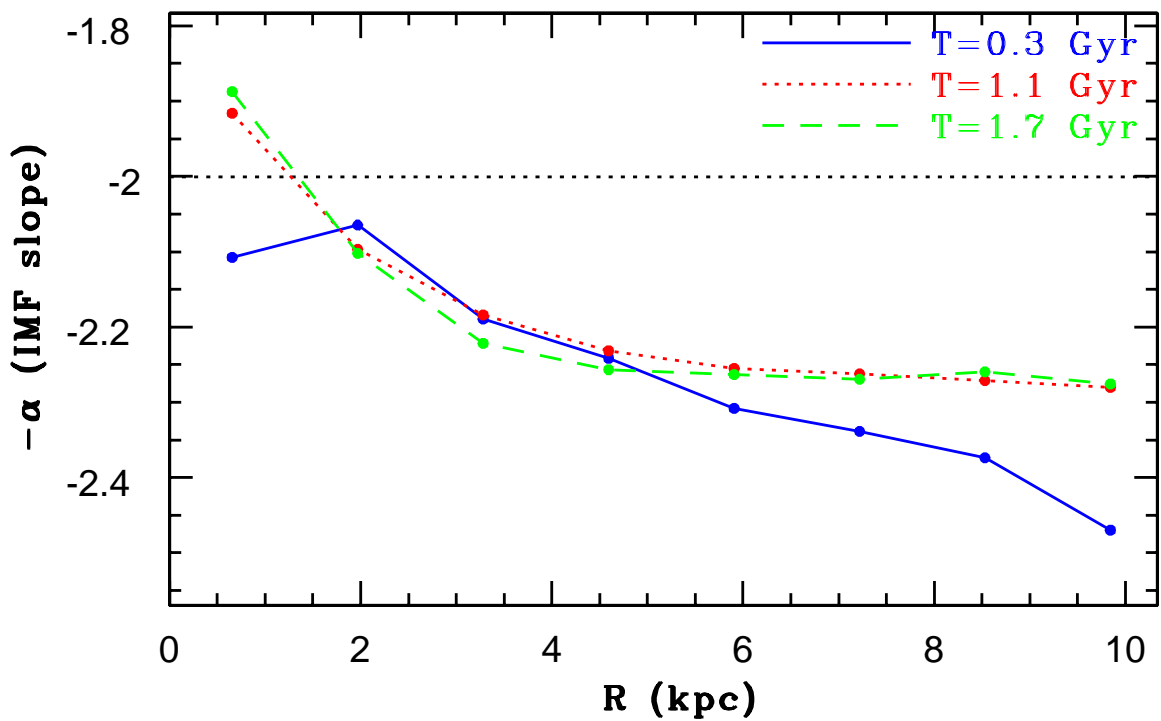}
\figcaption{
Radial profiles of $-\alpha$ for $R<10$ kpc at three epochs,
pre-starburst ($T=0.3$ Gyr; solid blue), during strong starburst ($T=1.1$ Gyr; dotted red),
and post-starburst ($T=1.7$ Gyr; dashed green) in the fiducial model.
The dotted black line 
indicates  the mean $\alpha$ for all new stars formed during the simulation.
The radial gradients are estimated by using all new stellar particles with $\alpha$.
The central 5 kpc at $T=1.1$ and 1.7 Gyr shows $d\alpha/dR \sim 0.07$ kpc$^{-1}$.
\label{fig-3}}
\end{figure}

\begin{figure}
\plotone{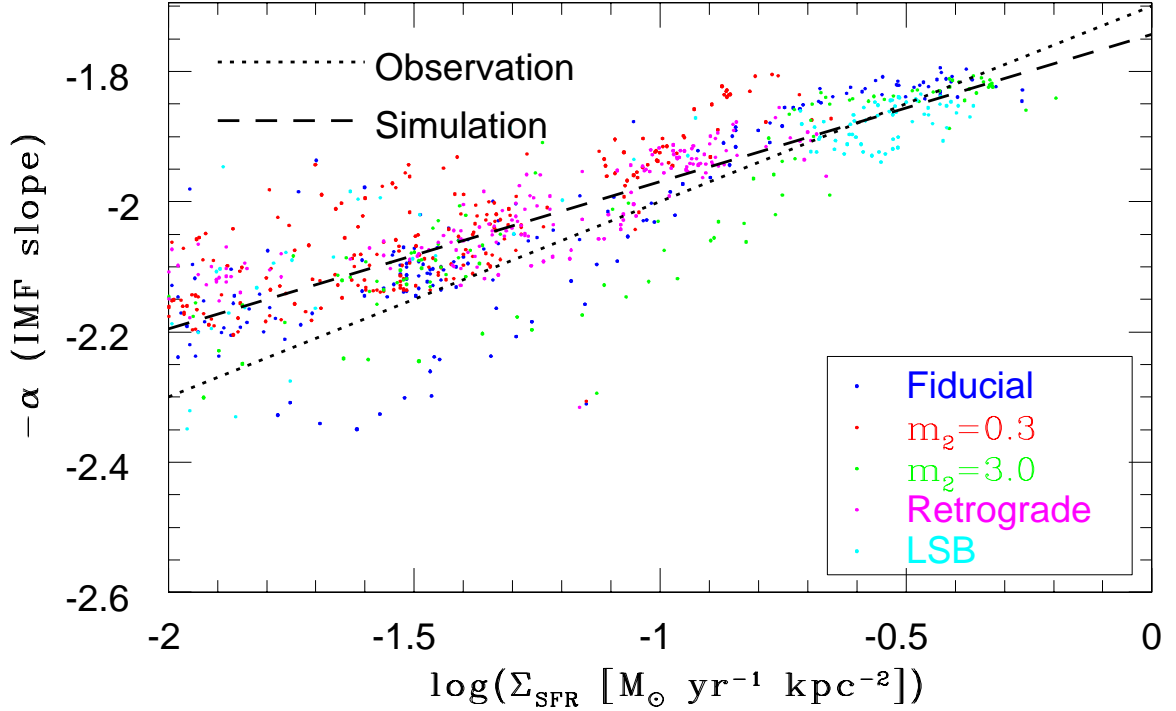}
\figcaption{
The locations of individual  star-forming regions on the $\log \Sigma_{\rm SFR}-\alpha$
plane. Different colors shows star-forming regions in different models: fiducial (blue),
weaker interaction with $m_2=0.3$ (red),  stronger interaction with $m_2=3$ (green), 
retrograde orbital configuration (magenta), and LSB model (cyan).
The observed $\alpha-\Sigma_{\rm SFR}$ relation by G11
($\alpha \approx -0.3 \log \langle \Sigma_{\rm SFR} \rangle +1.7$)
and the best-fit simulated one for
all of the  star-forming regions  are shown
by dotted and dashed black lines, respectively.
\label{fig-4}}
\end{figure}

\begin{figure}
\plotone{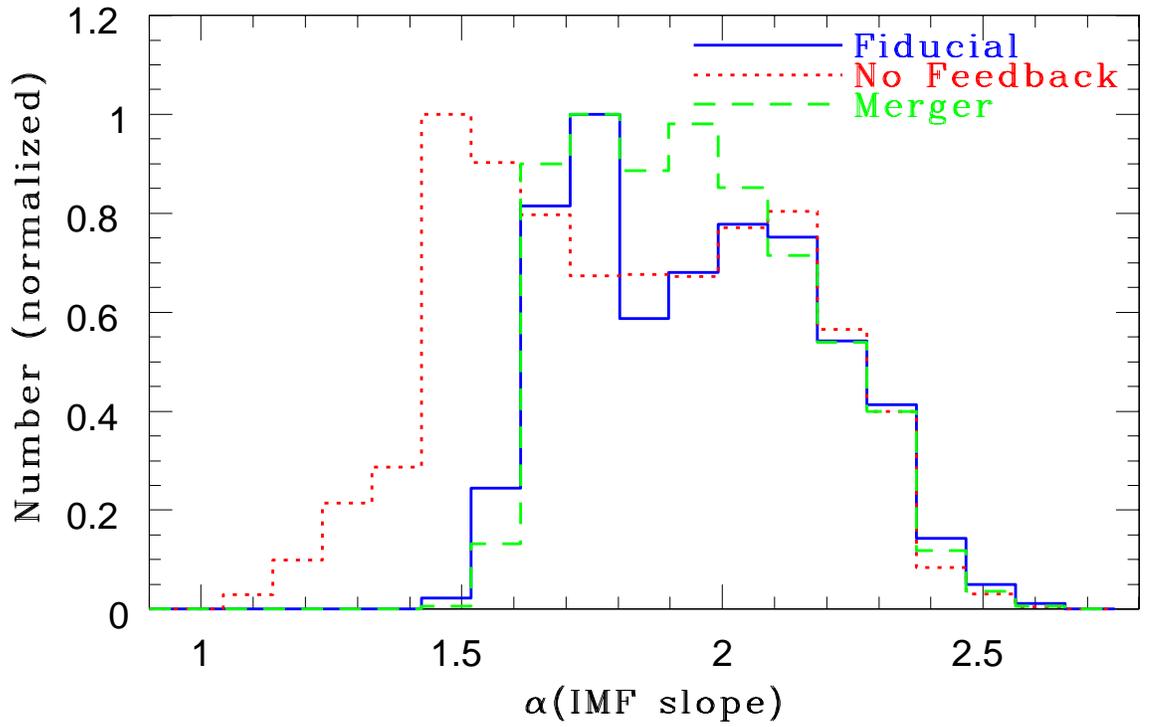}
\figcaption{
The number distributions of $\alpha$ for the fiducial model (solid blue),
the model with no SN feedback effects (dotted red), 
and the merger model (dashed green). The $\alpha$ distributions are derived
by using $\alpha$ of  all new stars in the models.
\label{fig-5}}
\end{figure}

\end{document}